# Adsorption and Dissociation of Toxic Gas Molecules on Graphene-like BC$_3$: A Search for Highly Sensitive Molecular Sensors and Catalysts

S. M. Aghaei[*] M. M. Monshi, I. Torres, and I. Calizo

Quantum Electronic Structures Technology Lab, Department of Electrical and Computer Engineering, Florida International University, Miami, Florida 33174, United States

*E-mail: smehd002@fiu.edu

**Abstract**

The adsorption behavior of toxic gas molecules (NO, CO, NO$_2$, and NH$_3$) on graphene-like BC$_3$ are investigated using first-principle density functional theory (DFT). The most stable adsorption configurations, adsorption energies, binding distances, charge transfers, electronic band structures, and the conductance modulations are calculated to deeply understand the impacts of the molecules above on the electronic and transport properties of the BC$_3$ monolayer. The graphene-like BC$_3$ monolayer is a semiconductor with a band gap of 0.733 eV. The semi-metal graphene has a low sensitivity to the abovementioned molecules. However, it is discovered that all the above gas molecules are chemically adsorbed on the BC$_3$ sheet with the adsorption energies less than −1 eV. The NO$_2$ gas molecule is totally dissociated into NO and O species through the adsorption process, while the other gas molecules retain their molecular forms. The amounts of charge transfer upon adsorption of CO and NH$_3$ gas molecules on BC$_3$ are found to be small. Hence, the band gap changes in BC$_3$ as a result of interactions with CO and NH$_3$ are only 4.63% and 16.7%, indicating that the BC$_3$-based sensor has a low and moderate sensitivity to CO and NH$_3$, respectively. Contrariwise, upon adsorption of NO or NO$_2$ on BC$_3$, a significant charge is transferred from the molecules to the BC$_3$ sheet, causing a semiconductor-metal transition. It is found that the BC$_3$-based sensor has high potential for NO detection due to the significant conductance changes, moderate adsorption energy, and short recovery time. More excitingly, the BC$_3$ is a likely catalyst for dissociation of the NO$_2$ gas molecule. Our findings divulge promising potential of the graphene-like BC$_3$ as a highly sensitive molecular sensor for NO and NH$_3$ detection and a catalyst for NO$_2$ dissociation.

1. Introduction

The need for miniaturized sensors with high sensitivity, fast response, high selectivity, high reliability, quick recovery, and low cost has motivated the scientists to seek new gas sensing systems based on novel nanomaterials. Graphene, the first discovered two-dimensional (2D) atomic crystal [1, 2], has enticed great interest thanks to its extraordinary properties, for instance, high surface-volume ratio, high carrier mobility, high chemical stability, low electronic temperature noise, high thermal stability, and fast response time. Ergo, it offers promise in the development of ultrasensitive gas sensors with high selectivity, fast recovery, high packing density, and low power consumption [3, 4]. The applicability of graphene in the field of gas sensing has been widely investigated both experimentally [5-7] and theoretically [8-12]. Pristine graphene shows low sensitivity toward common gas molecules such as CO, $CO_2$, $CH_4$, $N_2$, $NO_2$, $NH_3$, $H_2$, and $H_2O$ [8, 9, 11] which limits its potential for detection of individual gas molecules [12]. It has been reported that functionalization, introducing dopants, and defects can tune the electronic and magnetic properties of the various nanomaterials [13-19]. It has been reported that the sensitivity of graphene-based gas sensors can be significantly improved by introducing the dopants or defects [9, 11, 18, 20-24]. Zhang *et al.* discovered strong interactions between B-doped, N-doped, and defective graphene with small gas molecules such as $NO_2$, CO, NO, and $NH_3$ [9]. B-, N-, and Si-doped graphene indicated enhanced interactions with common gases such as $N_2$, NO, $NO_2$, $NH_3$, $SO_2$, CO, $CO_2$, $O_2$, $H_2$, and $H_2O$ compared to pristine graphene [11]. In another study, graphene doped with transition metals (Fe, Co, Ni, Ru, Rh, Pd, Os, Ir, and Pt) exhibited high sensitivity toward $O_2$ adsorption [21]. Borisova *et al.* discovered that the interactions of $H_2S$ with C atoms of defected graphene are much stronger than those of pristine graphene [25]. Density functional theory (DFT) calculations on Eu decorated single- and double-sided graphene sheets showed that each Eu could firmly bind to six hydrogen molecules [22].

Inspired by the astonishing gas sensing performance of graphene, the sensing capability of other 2D structures such as $MoS_2$ [26, 27], $WS_2$ [28, 29], phosphorene [30, 31], boron nitride [32, 33], silicene [34, 35], and germanene [36] toward various gases have been investigated. Recently, the graphene-like $BC_3$ sheet has been epitaxially grown on the $NbB_2$ (0001) surface [37]. The 2D honeycomb structures of $BC_3$ and graphene are analogous because of the similar atomic radius of B (87 pm) and C (67 pm). The B atoms are orderly distributed in the sheet and each B atom binds to three C atoms so that six C atoms from a hexagon surrounded by six B atoms [38, 39]. Unlike the zero-gap semi-metal graphene, $BC_3$ is a semiconductor with a band gap of 0.46-0.73 eV [38-43]. It was shown that the pristine [44], Li-doped [45], polylithiated molecule-doped [46], and transition metal-doped [47-49] $BC_3$ have high potential for $H_2$ storage. The capability of $BC_3$ as a gas sensor has been evaluated in the literature [50-53]. Beheshtian *et. al.* investigated the electronic sensitivity of pristine, Al-, and Si-doped $BC_3$ sheets to formaldehyde ($H_2CO$) molecule using DFT [50]. They found that although $H_2CO$ is weakly adsorbed on the sheet, both Al and Si doping enhance the reactivity of the $BC_3$ sheet toward $H_2CO$. Peyghan *et. al.* introduced $BC_3$ nanotubes as a potential gas sensor for CO detection [51].

In this work, first-principles methods based on DFT are employed to study the electronic and transport properties of the graphene-like $BC_3$ sheet after interaction with toxic gas molecules such as NO, $NO_2$, $NH_3$, and CO. The sensitivity of the $BC_3$-based sensors are evaluated from the variations in their electronic transport properties. Our results reveal the promising future of graphene-like $BC_3$ in the development of ultrahigh sensitive gas sensors.

## 2. Computation details

The results presented in this study are obtained using first-principle DFT calculations implemented in Atomistix ToolKit (ATK) package [54-56]. The Generalized Gradient Approximation of Perdew-Burke-Ernzerhof (GGA-PBE) exchange-correlation functional with a double-$\zeta$ polarized basis is adopted. The Grimme van der Waals (vdW) correction (PBE-D2) [57] is also engaged to describe long-range vdW interactions [58]. Furthermore, counterpoise correction (cp) is considered in the adsorption energy calculations to expunge the basis set superposition errors (BSSE) that arise due to the incompleteness of the linear combination of atomic orbitals (LCAO) basis set [59]. The energy mesh cut-off and the electronic temperature are set to be 150 Rydberg and 300°K, respectively. The supercell geometry with periodic boundary condition is accepted, and a large vacuum of 20 Å is considered in x direction (in which the structure is not periodic) to avoid the image-image interactions. All the structures are allowed to fully relax using the conjugate gradient method until the force on each atom is less than 0.01 eV/Å. The first Brillouin zones are sampled using 1×11×11 and 1×21×21 *k*-points for optimization and calculations, respectively.

The adsorption energy of gas molecules on the $BC_3$ sheet is calculated by:

$$E_{ad} = E_{BC_3+Molecule} - E_{BC_3} - E_{Molecule} \qquad (1)$$

where $E_{BC_3+Molecule}$, $E_{BC_3}$, and $E_{Molecule}$ are the total energies of the BC$_3$-Molecule complex, pristine BC$_3$ sheet, and the isolated gas molecule, respectively. The charge transfer upon adsorption of gas molecules on the BC$_3$ sheet is studied using Mulliken population analysis from the differences in the charge concentrations before and after adsorption.

To study the transport properties, the gas sensing system is divided into three regions: the central region (scattering region), left, and right electrodes, as shown in Fig. 1. The conductance of the system can be described using transmission coefficient at the Fermi level:

$$C(\varepsilon) = G_0 T(\varepsilon) \qquad (2)$$

where $G_0 = 2e^2/h$ is the quantum conductance, in which $e$ is the electron charge and $h$ is Planck's constant. Furthermore, transmission coefficients obtained from the retarded Green's function $G(\varepsilon)$ are:

$$T(\varepsilon) = G(\varepsilon)\Gamma^L(\varepsilon)G^\dagger(\varepsilon)\Gamma^R(\varepsilon) \qquad (3)$$

Here, $\Gamma^{(L)R}$ is the broadening function of the left (right) electrode which is:

$$\Gamma^{L(R)} = \frac{1}{i}\left(\sum\nolimits^{L(R)} - (\sum\nolimits^{L(R)})^\dagger\right) \qquad (4)$$

In addition, $\Sigma^{L(R)}$ is the electrode self-energy of the left (right) electrode.

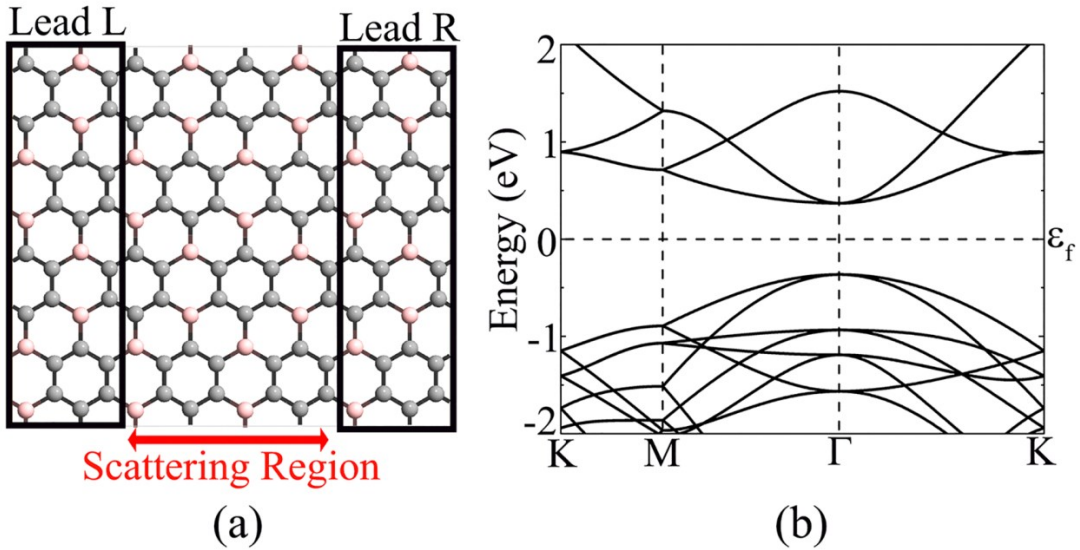

Fig. 1. (a) Schematic structural model of BC$_3$ gas sensor with two electrodes (black boxes). The grey and pink balls represent C and B atoms, respectively. (b) The band structure of a pristine BC$_3$ sheet. BC$_3$ is a semiconductor with a band gap of 0.733 eV.

## 3. Results and discussion

*3.1 Adsorption configurations*

We first test the accuracy of our computational method by calculating the band structure of a pristine $BC_3$ sheet, as shown in Fig. 1(b), and comparing our result with those reported previously. Our calculations show that $BC_3$ is a semiconductor with a band gap of 0.733 eV. The conduction band minimum (CBM) moved to the Γ-point due to the zone folding. Moreover, the B-C and C-C bond lengths are 1.57 and 1.43 Å, respectively. These results are in good agreement with literature data [38-43, 60].

In a $BC_3$ sheet, each six C atoms form a hexagon, and each B atom is attached to three separate C hexagons. Because the gas molecules tend to be adsorbed in various configurations, a number of input geometries should be taken into account. To this end, a single gas molecule of NO, $NO_2$, CO, and $NH_3$ is placed in a distance of 2 Å above C atom, B atom, C-C bond, B-C bond, center of C hexagons, and center of B-C hexagons. At these specific positions, several molecular orientations could be considered. On the hand, for diatomic molecules (NO and CO), the molecular axis could be aligned in parallel or perpendicular with respect to the surface of the $BC_3$ sheet. Moreover, the O atom of them can point up or down. On the other hand, two initial orientations for the triatomic ($NO_2$) and tetratomic ($NH_3$) molecules are considered. In the first orientation, the N atom of the molecules points toward the $BC_3$, while in the second one, the O atoms of $NO_2$ and the H atoms of $NH_3$ point down to the $BC_3$ sheet. After that, all the structures are allowed to fully relax. The interactions of the molecules with $BC_3$ sheet can be described regarding their adsorption energies. Based on Equation 1, the more negative the value of $E_{ad}$ is, the stronger adsorption of gas molecules on $BC_3$ would be. The most energetically favorable adsorption configurations are selected for further studies. Figure 2 presents the lowest energy configurations among all considered arrangements. The adsorption energies of aforementioned gas molecules on $BC_3$ sheet calculated at PBE-D2 level of DFT, the binding distances, and the net charge transfers using Mulliken population analysis are listed in Table 1.

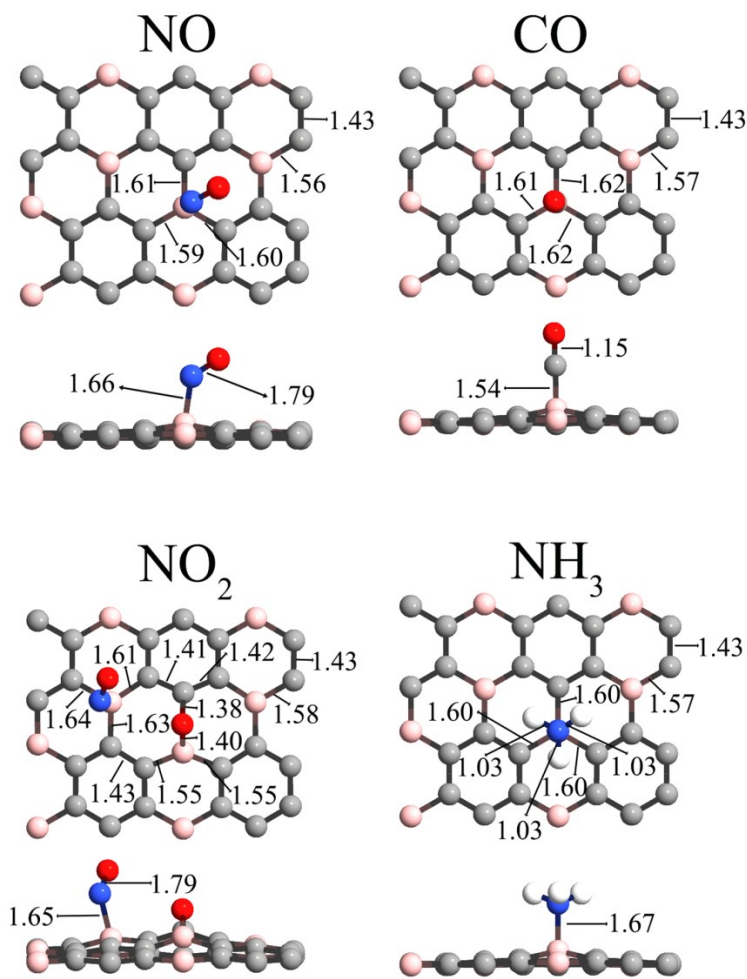

Fig. 2. The most stable adsorption configurations (top and side view) for NO, CO, NO$_2$, and NH$_3$ on the pristine BC$_3$ sheet. The grey, pink, blue, red, and white balls represent C, B, N, O, and atoms, respectively. The bond lengths (in Å) and the binding distances between the molecules and the BC$_3$ sheet are also given.

*3.2 NO Adsorption on BC$_3$ Sheet*

Upon exposure NO to BC$_3$ sheet, NO adopts a tilted orientation with respect to the plane of BC$_3$, as shown in Fig. 2. The N atom of NO is pointing to B atom of BC$_3$. The O-N-B angle is 129.51°. The interaction between the electron-deficient B atom and the electron-donating N atom of the NO leads to a moderate adsorption energy of −1.11 eV and the formation of a tight N-B bond (1.66 Å), suggesting that NO is chemically adsorbed on the BC$_3$ sheet. It should be noted that the B-N distance (1.66 Å) in NO-BC$_3$ complex is almost identical to the bond length of B-N in ammonia borane ($BH_3NH_3$) (1.6576 Å) [61], proving the formation of a covalent bond between $BC_3$ and NO. The BSSE causes an artificial attraction between the molecule and its adsorbents which leads to a strong binding. To exclude the BSSE, the Bernardi counterpoise correction is considered [59]. The adsorption energies with counterpoise correction for the different systems are tabulated in Table 1. The absolute value for NO adsorption on $BC_3$ is decreased from 1.11 to 0.91 eV, respectively. The obtained adsorption energy for NO on $BC_3$ is significantly higher compared to those reported for its adsorption on pristine graphene (−0.30 eV) and N-doped graphene (−0.40 eV) [9]. It is also comparable to the reported adsorption energy of NO on B-doped graphene (−1.07 eV) [9].

The $BC_3$ structure undergoes a structural change upon NO adsorption. The bond angles between N-B-C are 92.42°, 95.20°, and 108.42°. The C-B bond length is elongated from 1.57 Å to 1.59-1.61 Å around the interaction area. Moreover, the B site is transformed from $sp^2$ hybridization to $sp^3$ hybridization with an average buckling distance of ~0.242 Å. Similar observations have been reported for B-doped graphene [9] and B-doped carbon nanotubes [62]. Furthermore, the N-O bond length is increased from 1.173 Å in an isolated NO molecule to 1.179 Å in the NO-$BC_3$ complex. The small difference between the bond length values of NO molecule before and after adsorption on $BC_3$ shows that there is no dissociative adsorption of NO on $BC_3$. When NO is adsorbed on $BC_3$ sheet, an apparent charge of 0.65 $e$ is transferred from NO to the $BC_3$ sheet. The N and O atom experience a decrease of 0.52 and 0.13 $e$ in their charge states, respectively.

*3.3 CO Adsorption on $BC_3$ Sheet*

Next, the adsorption mechanism of the CO gas molecule on $BC_3$ sheet is studied. The most stable adsorption configuration of CO-$BC_3$ complex is shown in Fig. 2. The CO molecule is placed perpendicular to the $BC_3$ plane with the C atom at B atom and the other way around. The values of adsorption energy without and with counterpoise correction are −1.34 and −1.12 eV, respectively. The minimum atom-atom distance between the CO and $BC_3$ (C-B bond length) is 1.54 Å which is so close to the C-B dimer bond length (1.56 Å). These results reveal that the CO is chemically adsorbed on the $BC_3$ sheet. The calculated $E_{ad}$ of CO on $BC_3$ is significantly higher than values reported for CO adsorption on pristine graphene (−0.12 eV), N-doped graphene (−0.40 eV), and B-doped graphene (−0.14 eV) [9]. The structure of $BC_3$ is slightly altered in the interaction area. The B atom protrudes outwards after CO adsorption with an average buckling distance of ~0.280 Å. The bond angles between C-B-C are 99.26°, 99.59°, and 101.05°. The C-B bonds in $BC_3$ slightly extended to ~1.62 Å after CO adsorption in comparison with 1.57 Å in the pristine $BC_3$ sheet. The bond lengths of CO before and after interaction with CO are 1.14 and 1.15 Å, respectively, indicating that the CO molecule is not dissociated through the adsorption process. Moreover, the interaction between CO and $BC_3$ results in a charge transfer of 0.09 $e$ from $BC_3$ sheet to CO. The C and O atoms of the CO molecule loses and attains electronic charges of 0.05 and 0.14 $e$, respectively.

*3.4 $NO_2$ Adsorption on $BC_3$ Sheet*

The $NO_2$ adsorption mechanism on $BC_3$ is more complicated than the other molecules studied above. The $BC_3$ sheet and $NO_2$ molecule undergo significant structural changes upon adsorption on $BC_3$. The N-O bond lengths in an isolated $NO_2$ (1.21 Å) get extended to 1.79 and 3.11 Å, indicating that the molecule is fully dissociated during the adsorption process into the chemisorbed NO and O species. The N-O-N bond angle of an isolated $NO_2$ decreases from 133.06° to 95.04° after complexation with $BC_3$. The O atom of the molecule forms two covalent bonds with C and B atoms of the $BC_3$ sheet, where the O-C and O-B bond lengths are 1.38 and 1.40 Å, respectively. The C-O-B bond angle is 100.68°. On the other hand, the NO part of the $NO_2$ molecule binds to another B atom of the same B-C hexagon with a N-B covalent bond. The chemisorption of the dissociated NO on $BC_3$ is quite similar to that of the NO molecule on the $BC_3$ sheet. The O-N-B angle is 130.15°, and the N-B bond length is 1.65 Å. The adsorption energy of $NO_2$ on $BC_3$ sheet without and with counterpoise correction are −1.69 and −1.41 eV, respectively. The high adsorption energy and short binding distance confirm the chemisorption/dissociation of $NO_2$ after interaction with a $BC_3$ sheet. The value of adsorption energy for $NO_2$-$BC_3$ is greater than those calculated for NO and CO molecules. More interestingly, the calculated $E_{ad}$ of $NO_2$ on $BC_3$ is greater than those reported for $NO_2$ adsorption on pristine graphene (−0.48 eV), N-doped graphene (−0.98 eV), and B-doped graphene (−1.37 eV) [9]. Moreover, a significant charge of 0.77 $e$ is transferred from $NO_2$ molecule to $BC_3$ sheet after complexation. This large charge transfer correlates to the strong adsorption energies of the $NO_2$ on $BC_3$. The first dissociated part of the $NO_2$ (O atom) loses electronic charge of 0.14 $e$. The second dissociated part of the $NO_2$ (NO) loses 0.63 $e$, which is comparable to the obtained charge transfer for NO-BC$_3$ complex (0.65 $e$). The charge states on N and O are reduced by 0.54 and 0.09 $e$, correspondingly.

*3.5 NH$_3$ Adsorption on BC$_3$ Sheet*

Afterward, the interactions between the NH$_3$ molecule and BC$_3$ sheet are investigated. It is found that NH$_3$ is bound to the B atom with the N atom pointing at the sheet, as shown in Fig. 2. The N-B binding distance is 1.67 Å which is close to the B-N bond length in BH$_3$NH$_3$ (1.6575 Å) [61], indicating the formation of covalent bond between NH$_3$ and BC$_3$ sheet. The bond angles between N-B-C are 100.88°, 101.66°, and 101.74° The NH$_3$ is chemisorbed on the BC$_3$ sheet with an adsorption energy of −1.45 (−1.26) eV without (with) counterpoise correction. The calculated $E_{ad}$ for NH$_3$-BC$_3$ complex is significantly greater than those obtained for NH$_3$ adsorption on graphene (−0.11 eV), N-doped graphene (−0.12 eV), and B-doped graphene (−0.50 eV) [9]. It should be noted that this relatively high adsorption energy for the NH$_3$-BC$_3$ complex is accompanied by a small charge transfer of 0.10 $e$ from the molecule to BC$_3$ sheet which hinders its capability toward NH$_3$ detection. The N atom of NH$_3$ donates 0.1 $e$ to the BC$_3$ sheet through the adsorption process, while the charges on H atoms keep constant. The adsorption of NH$_3$ on BC$_3$ sheet brings about some structural changes in both the molecule and the BC$_3$ sheet. The B atom which is bound to N atom experiences a transformation from sp$^2$ to sp$^3$ hybridization where it stands out at an average distance of 0.317 Å above the planar sheet of BC$_3$. The B-C bonds around the interaction areas get extended to 1.60 Å. The bond lengths of N-H in an isolated NH$_3$ (1.03 Å) is found to be the same after adsorption on the BC$_3$ sheet. However, the H-N-H bond angles are increased from 104.58° in an isolated NH$_3$ to 108.16° in the NH$_3$-BC$_3$ complex.

Table 1. The calculated adsorption energy ($E_{ad}$) with and without counterpoise correction, binding distance which is the shortest atom to atom distance between molecule and the $BC_3$ sheet (D), and the charge transfer (Q), energy band gap ($E_g$), energy band gap changes ($\Delta E_g$), and recovery time ($\tau$). The negative values of charge indicate a charge transfer from molecule to the $BC_3$ sheet.

| System | $E_{ad}$ (eV) | $E_{ad\ C.P}$ (eV) | D (Å) | Q (e) | $E_g$ (eV) | $\Delta E_g$ (%) | $\tau$ (sec) |
|---|---|---|---|---|---|---|---|
| Pure $BC_3$ | - | - | - | - | 0.733 | - | - |
| $BC_3$-NO | − 1.11 | − 0.91 | 1.66 | −0.65 | 0.000 | 100 | 0.02 |
| $BC_3$-CO | − 1.34 | − 1.12 | 1.54 | +0.09 | 0.767 | 4.63 | 11.75 |
| $BC_3$-$NO_2$ | − 1.69 | − 1.41 | 1.38 | −0.77 | 0.000 | 100 | 174662 |
| $BC_3$-$NH_3$ | − 1.45 | − 1.26 | 1.67 | −0.10 | 0.610 | 16.7 | 240.6 |

*3.5 Electronic and transport properties of toxic gas molecules-$BC_3$ complexes*

To gain more insight into the adsorption of gas molecules on a $BC_3$ sheet, their electronic total charge densities are calculated. As can be seen in Fig. 3, an orbital overlap can be observed between NO, CO, $NO_2$, and $NH_3$ gas molecules and the $BC_3$ sheet, revealing the occurrence of a strong chemisorption/dissociation. These results are in accordance with obtained adsorption energies and binding distances.

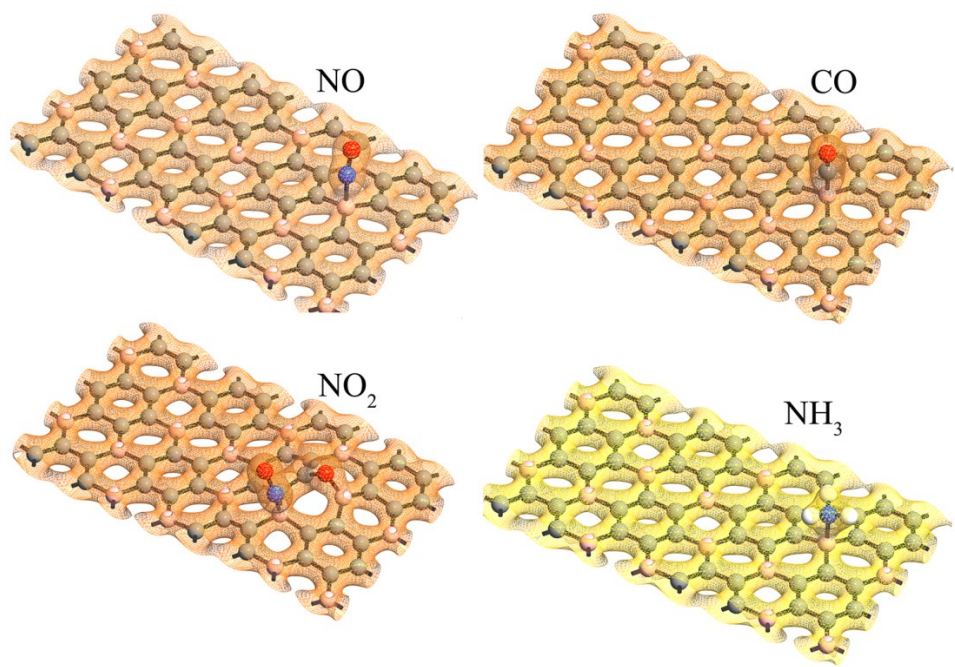

Fig. 3. The electronic total charge densities for the adsorption of NO, CO, $NO_2$, and $NH_3$ gas molecules on pristine $BC_3$.

The orbital mixing and the charge transfer are expected to bring significant changes to the electronic structure of the $BC_3$ sheet which is beneficial for sensing applications. The band gap can determine the electrical conductivity ($\sigma$) of materials as follow [63]:

$$\sigma \propto \exp(-E_g / 2k_B T) \qquad (5)$$

Here, T is the temperature and $k_B$ is the Boltzmann constant. Apparently, a small change in the energy band gap dramatically alters the electrical conductivity. Figure 4 presents the band structures of $BC_3$ after complexation with NO, CO, $NO_2$, and $NH_3$. The energy band gap of $BC_3$ (0.733 eV) increases to 0.767 eV upon interaction with CO and decreases to 0.610 eV after interaction with $NH_3$. The band gap changes ($\Delta E_g$) for the CO-$BC_3$ complex is 4.63% and for the $NH_3$-$BC_3$ complex is 16.7%. These small variations in the band gap of $BC_3$ after interaction with CO and $NH_3$ agree well with the small charge transfer of CO-$BC_3$ (+0.09 $e$) and $NH_3$-$BC_3$ (−0.10 $e$) systems. As a result, the electrical conductivity slightly decreases and slightly increases upon interaction with CO and $NH_3$ gas molecules, respectively. Although the values of adsorption energy of CO (−1.12 eV) and $NH_3$ (−1.26 eV) on $BC_3$ are relatively high, the values of $\Delta E_g$ imply that $BC_3$-based gas sensor has a low sensitivity to CO and a moderate sensitivity to the $NH_3$ molecule.

In contrast, the influence of NO and $NO_2$ adsorption on the electronic structure of $BC_3$ is much more pronounced. As can be seen in Fig. 4, there are almost flat bands on the Fermi levels of NO- and $NO_2$-$BC_3$ systems, indicating that the systems are metallic. One possible reason for this phenomena could be the occurrence of large transfer after adsorption of NO (−0.65 $e$) and $NO_2$ (−0.77 $e$) molecules on $BC_3$. It should be noted that the alterations in the electronic structure of $BC_3$ upon interaction with $NO_2$ are more noticeable compared to interaction with NO. These results can be confirmed by the obtained adsorption energies. A significant increase (100%) in the electrical conductivity of $BC_3$-based sensor upon adsorption of NO ($NO_2$) is expected, confirming its high sensitivity toward NO ($NO_2$) adsorption (dissociation).

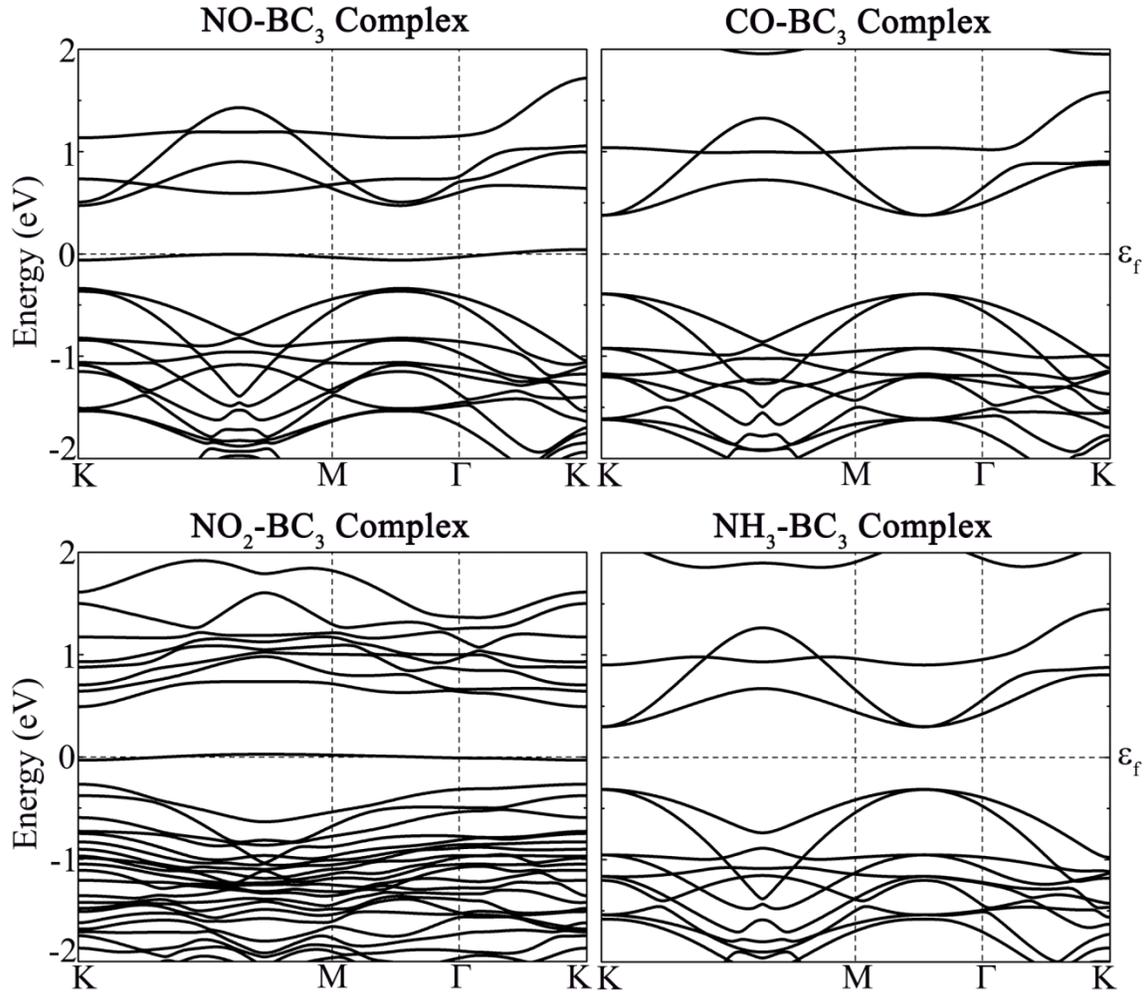

Fig. 4. Band structures of BC$_3$ sheet after interaction with NO, CO, NO$_2$, and NH$_3$ gas molecules

An important factor for the evaluation of the performance of a gas sensor is the recovery time of the gas sensing material. Novoselov *et al.* have shown that the graphene sensor could be recovered to its initial geometry by annealing at 150°C in a vacuum or ultraviolet (UV) irradiation within 100-200 s [64]. The recovery time $\tau$ can be expressed using the conventional transition state theory as follows:

$$\tau \propto \upsilon_0^{-1} \exp(-E_{ad}/k_B T) \qquad (6)$$

Here, $\upsilon_0$ is the attempt frequency. Based on this equation, a much longer recovery time is expected, if the adsorption energy is considerably increased (more negative). Based on our findings, the NO on BC$_3$ has the shortest recovery time followed by CO, NH$_3$, and NO$_2$. As an example, the recovery time of BC$_3$-based gas sensor after NO, CO, NH$_3$, and NO$_2$ adsorption could be ~ 0.02, 11.75, 174662, and 240.6 s by annealing at 150˚C in UV irradiation ($\upsilon_0$ ~ THz), see Table 1.

To further investigate the effects of gas molecules on the conductance of $BC_3$ sheet, the quantum conductances of the $BC_3$-based sensor (see Fig. 1(a)) before and after adsorption of gas molecules are calculated, as shown in Fig. 5. One can see that the values of conductance of CO-$BC_3$ and $NH_3$-$BC_3$ complexes slightly decrease and increase at low energies compared to that of pristine $BC_3$, respectively. However, the alteration of the conductance is more vivid for the later complex which concurs well with our previous results. More excitingly, large peaks appear at Fermi levels of NO-$BC_3$ and $NO_2$-$BC_3$ complexes which confirm the metallic behavior of these systems. The variations in the conductance of the $BC_3$ after $NO_2$ adsorption are more noticeable in comparison with that after NO adsorption. All these findings show that the $BC_3$-based sensor has great potential for NO detection due to the significant conductance changes, moderate adsorption energy, and short recovery time. Although, the sensitivity of $BC_3$ to CO gas molecule is low, this sensor could detect $NH_3$ molecule. More interestingly, the $BC_3$ might be a promising catalyst for dissociation of $NO_2$ gas molecule.

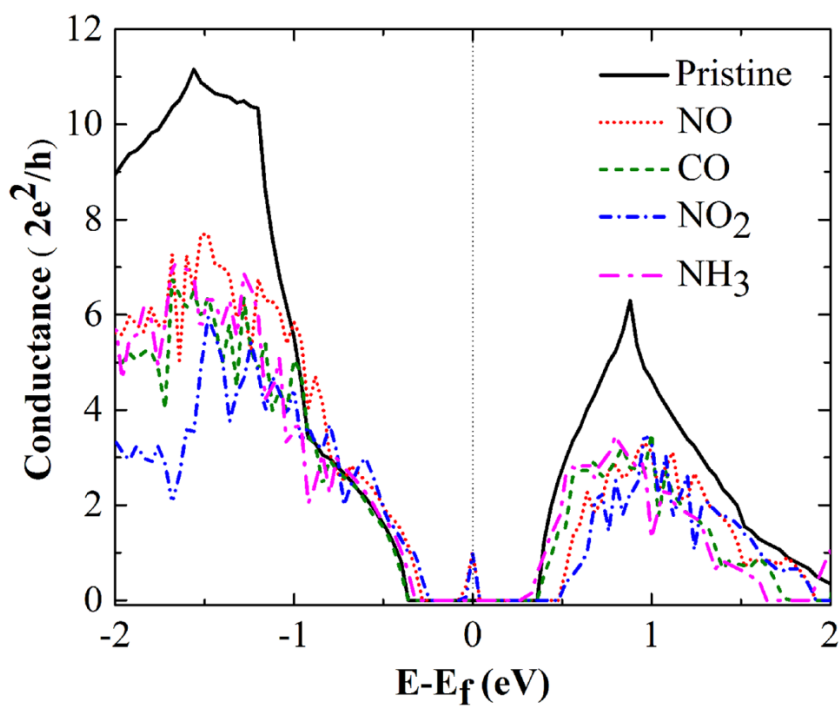

Fig. 5. Quantum conductance of $BC_3$-based sensor before and after NO, CO, $NO_2$, and $NH_3$ gas molecules adsorption.

## 4. Conclusions

We employed DFT to scrutinize the interaction of $BC_3$ with toxic gas molecules. While NO, CO, $NO_2$, and $NH_3$ gas molecules are weakly physisorbed on the surface of graphene, our results indicated that the presence of active B atoms on $BC_3$ can strengthen the adsorption of these molecules. Although CO, NO, and $NH_3$ tend to be adsorbed in the molecular form, $NO_2$ molecule is fully dissociated into chemisorbed NO and O species during adsorption process. The calculated band structures reveal that the band gap of $BC_3$ (0.733 eV) is increased 4.63% after CO adsorption and decreased 16.7% after $NH_3$ adsorption. These variations in the electronic properties of $BC_3$ correspond to small charge transfers from CO (−0.1 $e$) and $NH_3$ (+0.09 $e$) to the adsorbent. These findings suggest that the $BC_3$-based sensor has low and moderate sensitivities to CO and $NH_3$. Significant charge transfers happen upon adsorption of NO (−0.65 $e$) and $NO_2$ (−0.77 $e$) on $BC_3$ sheet, resulting in a semiconductor-metal transition in $BC_3$. These alternations in the band gap of $BC_3$ cause a significant increase in its conductance. We demonstrated that the $BC_3$-based sensor has a great potential for NO detection due to the significant conductance changes, moderate adsorption energy (−1.11 eV), and short recovery time (0.02 s). Interestingly, $BC_3$ might be a promising catalyst for dissociation of $NO_2$ gas molecule. On the basis of our theoretical findings, graphene-like $BC_3$ can be considered as a highly sensitive molecular sensor for NO and $NH_3$ and a potential catalyst for $NO_2$ dissociation.

**Acknowledgment**

This work was supported in part by the Florida Education Fund's McKnight Junior Faculty Fellowship.

## References

[1] K.S. Novoselov, A.K. Geim, S.V. Morozov, D. Jiang, Y. Zhang, S.V. Dubonos, I.V. Grigorieva, A.A. Firsov, Electric field effect in atomically thin carbon films, Science, 306 (2004) 666-669.

[2] K.S. Novoselov, A.K. Geim, S. Morozov, D. Jiang, M. Katsnelson, I. Grigorieva, S. Dubonos, A. Firsov, Two-dimensional gas of massless Dirac fermions in graphene, Nature, 438 (2005) 197-200.

[3] S. Basu, P. Bhattacharyya, Recent developments on graphene and graphene oxide based solid state gas sensors, Sens. Actuators B Chem., 173 (2012) 1-21.

[4] Q. He, S. Wu, Z. Yin, H. Zhang, Graphene-based electronic sensors, Chem. Sci., 3 (2012) 1764-1772.

[5] J.D. Fowler, M.J. Allen, V.C. Tung, Y. Yang, R.B. Kaner, B.H. Weiller, Practical chemical sensors from chemically derived graphene, ACS Nano, 3 (2009) 301-306.

[6] G. Ko, H.-Y. Kim, J. Ahn, Y.-M. Park, K.-Y. Lee, J. Kim, Graphene-based nitrogen dioxide gas sensors, Curr. Appl. Phys., 10 (2010) 1002-1004.

[7] G. Chen, T.M. Paronyan, A.R. Harutyunyan, Sub-ppt gas detection with pristine graphene, Appl. Phys. Lett., 101 (2012) 053119.

[8] O. Leenaerts, B. Partoens, F. Peeters, Adsorption of $H_2O$, $NH_3$, CO, $NO_2$, and NO on graphene: A first-principles study, Phys. Rev. B, 77 (2008) 125416.

[9] Y.-H. Zhang, Y.-B. Chen, K.-G. Zhou, C.-H. Liu, J. Zeng, H.-L. Zhang, Y. Peng, Improving gas sensing properties of graphene by introducing dopants and defects: a first-principles study, Nanotechnology, 20 (2009) 185504.

[10] G. Lu, L.E. Ocola, J. Chen, Reduced graphene oxide for room-temperature gas sensors, Nanotechnology, 20 (2009) 445502.

[11] J. Dai, J. Yuan, P. Giannozzi, Gas adsorption on graphene doped with B, N, Al, and S: a theoretical study, Appl. Phys. Lett., 95 (2009) 232105.

[12] J.T. Robinson, F.K. Perkins, E.S. Snow, Z. Wei, P.E. Sheehan, Reduced graphene oxide molecular sensors, Nano Lett., 8 (2008) 3137-3140.

[13] M. Topsakal, E. Aktürk, H. Sevinçli, S. Ciraci, First-principles approach to monitoring the band gap and magnetic state of a graphene nanoribbon via its vacancies, Phys. Rev. B, 78 (2008) 235435.

[14] A. Lopez-Bezanilla, W. Zhou, J.-C. Idrobo, Electronic and Quantum Transport Properties of Atomically Identified Si Point Defects in Graphene, J. Phys. Chem.Lett., 5 (2014) 1711-1718.

[15] S.M. Aghaei, I. Calizo, Band gap tuning of armchair silicene nanoribbons using periodic hexagonal holes, J. Appl. Phys., 118 (2015) 104304.

[16] S.M. Aghaei, M.M. Monshi, I. Torres, I. Calizo, Edge functionalization and doping effects on the stability, electronic and magnetic properties of silicene nanoribbons, RSC Adv., 6 (2016) 17046-17058.

[17] M.M. Monshi, S.M. Aghaei, I. Calizo, Edge functionalized germanene nanoribbons: impact on electronic and magnetic properties, RSC Adv., 7 (2017) 18900-18908.

[18] A.S. Rad, First principles study of Al-doped graphene as nanostructure adsorbent for $NO_2$ and $N_2O$: DFT calculations, Appl. Surf. Sci., 357 (2015) 1217-1224.

[19] S. Gupta, H. Sabarou, Y. Zhong, P. Singh, Phase evolution and electrochemical performance of iron doped lanthanum strontium chromite in oxidizing and reducing atmosphere, Int. J. Hydrogen Energy, 42 (2017) 6262-6271.

[20] J. Dai, J. Yuan, Adsorption of molecular oxygen on doped graphene: atomic, electronic, and magnetic properties, Phys. Rev. B, 81 (2010) 165414.

[21] F. Nasehnia, M. Seifi, Adsorption of molecular oxygen on VIIIB transition metal-doped graphene: A DFT study, Mod. Phys. Lett. B, 28 (2014) 1450237.

[22] W. Liu, Y. Liu, R. Wang, L. Hao, D. Song, Z. Li, DFT study of hydrogen adsorption on Eu-decorated single-and double-sided graphene, Phys. Status Solidi B, 251 (2014) 229-234.

[23] L. Ma, J.-M. Zhang, K.-W. Xu, V. Ji, A first-principles study on gas sensing properties of graphene and Pd-doped graphene, Appl. Surf. Sci., 343 (2015) 121-127.

[24] B. Wanno, C. Tabtimsai, A DFT investigation of CO adsorption on VIIIB transition metal-doped graphene sheets, Superlattices Microstruct., 67 (2014) 110-117.

[25] D. Borisova, V. Antonov, A. Proykova, Hydrogen sulfide adsorption on a defective graphene, Int. J. Quantum Chem., 113 (2013) 786-791.

[26] Q. He, Z. Zeng, Z. Yin, H. Li, S. Wu, X. Huang, H. Zhang, Fabrication of Flexible $MoS_2$ Thin-Film Transistor Arrays for Practical Gas-Sensing Applications, Small, 8 (2012) 2994-2999.

[27] D.J. Late, Y.-K. Huang, B. Liu, J. Acharya, S.N. Shirodkar, J. Luo, A. Yan, D. Charles, U.V. Waghmare, V.P. Dravid, Sensing behavior of atomically thin-layered $MoS_2$ transistors, ACS Nano, 7 (2013) 4879-4891.

[28] N. Huo, S. Yang, Z. Wei, S.-S. Li, J.-B. Xia, J. Li, Photoresponsive and gas sensing field-effect transistors based on multilayer $WS_2$ nanoflakes, Sci. Rep., 4 (2014) 5209.

[29] M. O'Brien, K. Lee, R. Morrish, N.C. Berner, N. McEvoy, C.A. Wolden, G.S. Duesberg, Plasma assisted synthesis of $WS_2$ for gas sensing applications, Chem. Phys. Lett., 615 (2014) 6-10.

[30] L. Kou, T. Frauenheim, C. Chen, Phosphorene as a superior gas sensor: selective adsorption and distinct IV response, J. Phys. Chem. Lett., 5 (2014) 2675–2681.

[31] A.N. Abbas, B. Liu, L. Chen, Y. Ma, S. Cong, N. Aroonyadet, M. Köpf, T. Nilges, C. Zhou, Black phosphorus gas sensors, ACS Nano, 9 (2015) 5618-5624.

[32] A.A. Peyghan, M. Noei, S. Yourdkhani, Al-doped graphene-like BN nanosheet as a sensor for para-nitrophenol: DFT study, Superlattices Microstruct., 59 (2013) 115-122.

[33] F. Behmagham, E. Vessally, B. Massoumi, A. Hosseinian, L. Edjlali, A computational study on the $SO_2$ adsorption by the pristine, Al, and Si doped BN nanosheets, Superlattices Microstruct., 100 (2016) 350-357.

[34] R. Chandiramouli, A. Srivastava, V. Nagarajan, NO adsorption studies on silicene nanosheet: DFT investigation, Appl. Surf. Sci., 351 (2015) 662-672.

[35] S.M. Aghaei, M.M. Monshi, I. Calizo, A theoretical study of gas adsorption on silicene nanoribbons and its application in a highly sensitive molecule sensor, RSC Adv., 6 (2016) 94417-94428.

[36] V. Nagarajan, R. Chandiramouli, NO 2 adsorption behaviour on germanene nanosheet–A first-principles investigation, Superlattices Microstruct., 101 (2017) 160-171.

[37] H. Tanaka, Y. Kawamata, H. Simizu, T. Fujita, H. Yanagisawa, S. Otani, C. Oshima, Novel macroscopic $BC_3$ honeycomb sheet, Solid State Commun., 136 (2005) 22-25.

[38] Q. Wang, L.-Q. Chen, J.F. Annett, Stability and charge transfer of $C_3B$ ordered structures, Phys. Rev. B, 54 (1996) R2271.

[39] D. Tomanek, R.M. Wentzcovitch, S.G. Louie, M.L. Cohen, Calculation of electronic and structural properties of $BC_3$, Phys. Rev. B, 37 (1988) 3134.

[40] Y. Miyamoto, A. Rubio, S.G. Louie, M.L. Cohen, Electronic properties of tubule forms of hexagonal $BC_3$, Phys. Rev. B, 50 (1994) 18360.

[41] Y. Ding, Y. Wang, J. Ni, Electronic structures of $BC_3$ nanoribbons, Appl. Phys. Lett., 94 (2009) 073111.

[42] Y. Ding, Y. Wang, J. Ni, Structural, electronic, and magnetic properties of defects in the $BC_3$ sheet from first principles, J. Phys. Chem. C, 114 (2010) 12416-12421.

[43] S.-s. Li, C.-w. Zhang, W.-x. Ji, F. Li, P.-j. Wang, Tunable electronic properties induced by a defect-substrate in graphene/$BC_3$ heterobilayers, Phys. Chem. Chem. Phys., 16 (2014) 22861-22866.

[44] F.-C. Chuang, Z.-Q. Huang, W.-H. Lin, M.A. Albao, W.-S. Su, Structural and electronic properties of hydrogen adsorptions on $BC_3$ sheet and graphene: a comparative study, Nanotechnology, 22 (2011) 135703.

[45] Z. Yang, J. Ni, Li-doped $BC_3$ sheet for high-capacity hydrogen storage, Appl. Phys. Lett., 100 (2012) 183109.

[46] Y. Li, T. Hussain, A. De Sarkar, R. Ahuja, Hydrogen storage in polylithiated $BC_3$ monolayer sheet, Solid State Commun., 170 (2013) 39-43.

[47] T. Hussain, D.J. Searles, K. Takahashi, Reversible Hydrogen Uptake by BN and $BC_3$ Monolayers Functionalized with Small Fe Clusters: A Route to Effective Energy Storage, J. Phys. Chem. A., 120 (2016) 2009-2013.

[48] M. Chen, Y.-J. Zhao, J.-H. Liao, X.-B. Yang, Transition-metal dispersion on carbon-doped boron nitride nanostructures: Applications for high-capacity hydrogen storage, Phys. Rev. B, 86 (2012) 045459.

[49] S.R. Naqvi, T. Hussain, P. Panigrahi, W. Luo, R. Ahuja, Manipulating energy storage characteristics of ultrathin boron carbide monolayer under varied scandium doping, RSC Adv., 7 (2017) 8598-8605.

[50] J. Beheshtian, A.A. Peyghan, M. Noei, Sensing behavior of Al and Si doped $BC_3$ graphenes to formaldehyde, Sens. Actuators B Chem., 181 (2013) 829-834.

[51] A.A. Peyghan, S. Yourdkhani, M. Noei, Working mechanism of a $BC_3$ nanotube carbon monoxide gas sensor, Commun. Theor. Phys., 60 (2013) 113.

[52] A.A. Peyghan, H. Soleymanabadi, Adsorption of H2S at Stone–Wales defects of graphene-like $BC_3$: a computational study, Mol. Phys., 112 (2014) 2737-2745.

[53] M.S. Mahabal, M.D. Deshpande, T. Hussain, R. Ahuja, Sensing Characteristics of a Graphene-like Boron Carbide Monolayer towards Selected Toxic Gases, Chemphyschem, 16 (2015) 3511-3517.

[54] J. Taylor, H. Guo, J. Wang, Ab initio modeling of quantum transport properties of molecular electronic devices, Phys. Rev. B, 63 (2001) 245407.

[55] M. Brandbyge, J.-L. Mozos, P. Ordejón, J. Taylor, K. Stokbro, Density-functional method for nonequilibrium electron transport, Phys. Rev. B, 65 (2002) 165401.

[56] Atomistix ToolKit (ATK) QuantumWise Simulator [Online]. Available: http://www.quantumwise.com

[57] S. Grimme, Semiempirical GGA-type density functional constructed with a long-range dispersion correction, J. Comput. Chem., 27 (2006) 1787-1799.

[58] S. Grimme, C. Mück-Lichtenfeld, J. Antony, Noncovalent interactions between graphene sheets and in multishell (hyper) fullerenes, J. Phys. Chem. C, 111 (2007) 11199-11207.

[59] S.F. Boys, F.d. Bernardi, The calculation of small molecular interactions by the differences of separate total energies. Some procedures with reduced errors, Mol. Phys., 19 (1970) 553-566.

[60] E. Chigo-Anota, M.A. Alejandro, A.B. Hernández, J.S. Torres, M. Castro, Long range corrected-wPBE based analysis of the $H_2O$ adsorption on magnetic $BC_3$ nanosheets, RSC Adv., 6 (2016) 20409-20421.

[61] D.A. Dixon, M. Gutowski, Thermodynamic properties of molecular borane amines and the $[BH_4^-][NH_4^+]$ salt for chemical hydrogen storage systems from ab initio electronic structure theory, J. Phys. Chem. A., 109 (2005) 5129-5135.

[62] L. Bai, Z. Zhou, Computational study of B-or N-doped single-walled carbon nanotubes as $NH_3$ and $NO_2$ sensors, Carbon, 45 (2007) 2105-2110.

[63] S.S. Li, Semiconductor physical electronics, Springer Science & Business Media, 2012.

[64] F. Schedin, A. Geim, S. Morozov, E. Hill, P. Blake, M. Katsnelson, K. Novoselov, Detection of individual gas molecules adsorbed on graphene, Nature Mater., 6 (2007) 652-655.